\documentclass[modern]{aastex701}

\usepackage{graphicx}
\usepackage{amsmath}
\usepackage{placeins}
\usepackage{tikz, pgfplots}
\pgfplotsset{compat=1.18}
\usetikzlibrary{positioning}
\usetikzlibrary{angles}
\usepackage{txfonts}
\usepackage{nicefrac}

\newcommand{\brac}[1]{\left(#1\right)}
\newcommand{\bracc}[1]{\left[#1\right]}

\begin{document}

\title{Investigating polarization signatures from GRB models}

\author[orcid=0000-0003-1667-3392]{Pieter van der Merwe}
\affiliation{Center for Space Research,\\ North-West University,\\ Potchefstroom,\\ South-Africa}
\email[show]{pietervdmerwejnr@gmail.com}  

\author[orcid=0000-0002-8434-5692]{Markus B\"ottcher} 
\affiliation{Center for Space Research,\\ North-West University,\\ Potchefstroom,\\ South-Africa}
\email[show]{Markus.Bottcher@nwu.ac.za}

\published{February 17, 2026}
\accepted{January 15, 2026}
\revised{December 29, 2025}
\received{November 17, 2025}

\begin{abstract}
There is still much debate around the inner workings of the GRB prompt emission phase with many questions still left unanswered. Polarization signatures offer a promising new avenue to discriminate between the various GRB prompt emission models. The aim of this study is to estimate energy and time resolved polarization signatures resulting from Inverse Compton (IC) scattering for two specific GRB prompt emission models, namely the backscatter-dominated cork model by \citet{Vyas2} and a Compton drag model by \citet{Lazzati2}. In order to achieve this we apply an IC polarization Monte Carlo (MC) algorithm to those two GRB models in order to estimate the expected polarization signatures. For the backscatter-dominated cork model we find polarization signatures below $\sim 10$~\%, likely below the detection limits of current or near-future X-ray and $\gamma$-ray polarimeters. Our results for the Compton drag model indicates polarization results consistent with that found by \citet{Lazzati3}. Furthermore, we find some energy and time dependence of the estimated polarization, particularly for the polarization angle.

\end{abstract}



\section{Introduction}
Since the first detection of Gamma-ray bursts (GRBs) in 1967, the study of GRBs has been an active area of study. GRBs are primarily broken up into the prompt emission phase and the afterglow phase \citep{Zhang_2018}. The afterglow phase is the most well understood and is believed to be produced by the jet of the GRB interacting with the interstellar medium (ISM) to form an external shock accelerating relativistic electrons which produce radio through X-ray emmission
via the synchrotron mechanism. In contrast to this the prompt emission phase is not well understood with the primary emission mechanism still being a topic of much debate. It has become clear that relying on the spectral information from GRB observations and models alone is proving insufficient to discriminate between 
GRB models and that X-ray and $\gamma$-ray polarization 
may provide additional discriminating diagnostics.
To this extent we investigate the polarization signatures due to Comptonization for two GRB models, namely the Backscattering cork model \citep[][see section 3]{Vyas1,Vyas2} and a Compton drag model \citep[][see section 4]{Lazzati1,Lazzati2}. This is done using a GPU Monte Carlo code developed from the Monte Carlo Compton scattering formalism introduced by \citet{Matt1996} discussed in section 2. 

\section{The Monte Carlo code developed to run on GPUs}\label{sec:MCAlg}

In this section the numerical algorithm used is presented. The core part of the Monte Carlo IC scattering routine is based on the algorithm used in the Monte-Carlo Applications for Partially Polarized Inverse External-Compton Scattering (MAPPIES) code by \citet{Dreyer_2021} which in turn takes elements from \citet{Matt1996} and \citet{MonteCarloBible}. What follows is a brief overview of the algorithm.\\

\subsection{Overview}\label{subsec:overview}
The algorithm used performs various calculations in three different reference frames, namely the lab frame denoted with a subscript ``lab", the emission-region frame denoted with subscript ``em", and the rest frame of a scattering electron denoted by subscript ``e". We assume that the emission-region is relativistically expanding radially outward within an opening angle $\theta_{j}$ centred around the $z$-axis with bulk Lorentz factor $\Gamma_{\rm BL}$. The polar and azimuthal angles of the photon in the lab frame are denoted by $\theta_{\rm lab}$ and $\phi_{\rm lab}$, respectively. 

Here and for all following sections all angles are expressed in units of radians unless stated otherwise.

The algorithm can be broken into several different sections. First the energy $\epsilon_{\rm lab}$, position four-vector $\underline{r}_{\rm lab}=\brac{t,\vec{r}_{\rm lab}}$, and direction three-vector $\vec{D}_{\rm lab}=\brac{\sin\theta_{\rm lab}\cos\phi_{\rm lab},\sin\theta_{\rm lab}\sin\phi_{\rm lab},\cos\phi_{\rm lab}}$ of the photon is drawn in the lab frame in a manner 
dependent on the model used. 
A unit polarization three-vector $\vec{P}_{\rm em}$ is drawn perpendicular to the direction of the photon. Finally, we construct the photon wave four-vector  $\underline{k}_{\rm lab} = \nicefrac{\epsilon_{\rm lab}}{h c}\brac{1,\vec{D}_{\rm lab}}$, and unit polarization four-vector $\underline{p}_{\rm lab}=\brac{0,\vec{P}_{\rm lab}}$.

Next, $\underline{k}_{\rm lab}$ and $\underline{p}_{\rm lab}$ are transformed to a local segment of the emission region using standard Lorentz transformation equations for an arbitrary direction, 
\begin{equation}
\begin{split}
    a_0' &= \Gamma\brac{a_{0}-\vec{\beta}\cdot\vec{A}},\\
    \vec{A}'     &= \vec{A} + \frac{\Gamma^{2}}{1+\Gamma}\brac{\vec{\beta}\cdot\vec{A}}\vec{\beta}-\Gamma\vec{\beta}a_{0},\
\end{split}
\end{equation}
where $\underline{a}=\brac{a_{0},\vec{A}}$ is the vector to be transformed, $\Gamma$ is the Lorentz factor of the new frame, and $\vec{\beta}$ is the velocity of the new frame. 
When transforming the polarization unit four-vector $\underline{p}$, it is important to project the transformed four-vector onto the plane perpendicular to the direction of the photon in the new frame \citep{Connors}.

Once in the local emission frame, the direction of motion and Lorentz factor $\gamma$ of an electron in the emission region are drawn and the photon is Lorentz transformed into the electron rest frame.
IC scattering is then simulated in the electron rest frame as described in section \ref{subsec:SimICScat} below, after which we transform back to the lab frame using two successive transformations as described above.

Back in the lab frame, the linear Stokes parameters $Q$ and $U$ of the individual photons are calculated from the scattered polarization three-vector $\vec{P}^{\rm sc}_{\rm lab}$. This is done following the formalism of \citet{Matt1996}.

We perform a ``time tagging" in order to estimate the difference of arrival times of each of the photons to a given observer. To do this we consider an infinite plane at distance $R_{\rm obs}$ to the emission region. The plane has a normal $\hat{n}$ parallel to the observer direction and $\vec{r}_{unsc}$ represents the point on the plane where an unscattered photon emitted from the source would intersect the plane. We then calculate the time $\Delta t$ it takes the photon to travel from its final position to the plane by solving

\begin{equation}
    \Delta t =  \frac{\hat{n}\cdot\brac{\vec{r}_{unsc}-\vec{r}_{lab}}}{c\hat{n}\cdot\hat{D}_{lab}}, 
\end{equation}

taking the time $t_{\rm obs} = t_{\rm lab}+\Delta t$ as the time of the detection of the photon.       

Finally, the results from all of the photons can be binned in terms of either energy or time along a given viewing angle. With the photons binned, the total linear Stokes parameters of each bin are calculated as 

\begin{equation}
\begin{split}
    Q_{bin} &= \sum_{i = 1}^{N_{bin}}\frac{Q_i}{N_{bin}}\\
    U_{bin} &= \sum_{i = 1}^{N_{bin}}\frac{U_i}{N_{bin}}.\\
\end{split}
\end{equation}

With $Q_{bin}$ and $U_{bin}$ calculated for each bin, the final polarization degree $\Pi_{bin}$ and polarization angle $\chi_{bin}$ for each bin are calculated as 

\begin{equation}
\begin{split}
    \Pi_{bin} &= \sqrt{Q_{bin}^{2}+U_{bin}^{2}}\\
    \chi_{bin} &= \frac{1}{2}\arctan\frac{U_{bin}}{Q_{bin}}.
\end{split}
\end{equation}

\subsection{Drawing electrons}\label{subsec:electron}

In this section and all following sections random numbers denoted by $\xi$ are drawn from a uniform distribution, with $\xi \in \bracc{0,1}$.

The electrons are drawn from a Maxwell-J\"uttner distribution through the use of the Sobol rejection method popularised by \citet{MonteCarloBible}.

Let $T_{e}$ be the temperature of the electron distribution, $\eta$ the dimensionless momentum of the electron, and $\Theta_{e}=\nicefrac{kT_{e}}{m_{e}c^{2}}$, then if $\Theta_{e} \leq 0.29$, two random numbers $\xi_{1}$ and $\xi_{2}$ are drawn \citep{MonteCarloBible}. A temporary quantity $\xi_{temp} = -\nicefrac{3}{2}\ln\xi_{1}$ is calculated. Then if $\xi_{2}^{2}<0.151\brac{1+\Theta_{e}\xi_{temp}}^{2}\xi_{temp}\brac{2+\Theta_{e}\xi_{temp}}\xi_{1}$, we set $\eta=\bracc{\Theta_{e}\xi_{temp}\brac{2+\Theta_{e}\xi_{temp}}}^{\nicefrac{1}{2}}$, otherwise new values for $\xi_{1}$ and $\xi_{2}$ are drawn. 

If $\Theta_{e} > 0.29$ four random numbers $\xi_{1},\xi_{2},\xi_{3}$ and $\xi_{4}$ are drawn \citep{MonteCarloBible}. Two values $\Xi_{1}=-\Theta_{e}\ln\brac{\xi_{1}\xi_{2}\xi_{3}}$ and $\Xi_{2}=-\Theta_{e}\ln\brac{\xi_{1}\xi_{2}\xi_{3}\xi_{4}}$ are calculated and compared such that if $\Xi_{2}^{2}-\Xi_{1}^{2}>1$, we set $\eta=\Xi_{1}$, otherwise new values are drawn and the process is repeated. Finally we calculate the Lorentz factor of the electron $\gamma_{e}=\bracc{\eta^{2}+1}^{\nicefrac{1}{2}}$. 

The direction of motion of the electron is drawn randomly from an isotropic distribution.

\subsection{Simulating inverse Compton Scattering}\label{subsec:SimICScat}

The probability of a photon to undergo Compton scattering is determined by the polarization-averaged Klein-Nishina cross section,
\begin{equation}
\begin{aligned}
    \sigma_{KN} &= \frac{3}{4}\sigma_{T}\bracc{\frac{1+x_{e}}{x_{e}^{3}}\brac{\frac{2x_{e}\brac{1+x_{e}}}{1+2x_{e}}-\ln\brac{1+2x_{e}}}+\frac{\ln\brac{1+2x_{e}}}{2x_{e}}-\frac{1+3x_{e}}{\brac{1+2x_{e}}^{2}}},\\
\end{aligned}
\end{equation}
where $x_{e}=\nicefrac{\epsilon_{e}}{m_{e}c^{2}}$ is evaluated in the electron rest frame, and $\sigma_{T}$ is the Thomson cross section \citep{Matt1996,Dreyer_2021}. 

The energy of the scattered photon is calculated as 

\begin{equation}
    \epsilon_{e}^{sc} = \frac{\epsilon_{e}}{1 + x_{e} (1 - \cos\theta_{e}^{sc})},
\end{equation}

where $\theta_{e}^{sc}$ is determined by numerically solving the equation
\begin{equation}
\begin{aligned}
    \xi_{\theta} &= \frac{x_{e}\bracc{\frac{3}{2}+\mu_{e}^{sc}\brac{1-\frac{1}{2}\mu_{e}^{sc}}}+\frac{1}{3}\bracc{1+\brac{\mu_{e}^{sc}}^{3}}}{\frac{2}{3}+2x_{e}+x_{e}^{-1}\ln\brac{1+2x_{e}}}\\
    &- \frac{x_{e}^{-1}\bracc{\ln\brac{1+x_{e}\bracc{1-\mu_{e}^{sc}}}-\ln\brac{1+2x_{e}}}}{\frac{2}{3}+2x_{e}+x_{e}^{-1}\ln\brac{1+2x_{e}}}\\
\end{aligned}
\end{equation}
for $\theta_{e}^{sc}$, where $\mu_{e}^{sc} = \cos\theta_{e}^{sc}$ and $\xi_{\theta}$ is a random number. The scattered azimuthal angle is numerically calculated from  
\begin{equation}
\begin{aligned}
    \xi_{\phi} &= \frac{1}{2\pi}\bracc{\phi_{e}^{sc}-\frac{\sin^{2}\theta_{e}^{sc}\sin\phi_{e}^{sc}\cos\phi_{e}^{sc}}{\frac{x_{e}}{x_{e}^{sc}}+\frac{x_{e}^{sc}}{x_{e}}-\sin^{2}\theta_{e}^{sc}}}.\\
\end{aligned}
\end{equation}

Finally, the contribution of the photon to the overall polarization is determined by drawing a random number $\xi_{PD} \in \bracc{0,1}$ and comparing it to the polarization degree of the scattered photon, $PD^{sc}_{e}$, which is calculated as
\begin{equation}
    PD^{sc}_{e} = 2\bracc{\frac{1-\sin^{2}\theta_{e}^{sc}\cos^{2}\phi_{e}^{sc}}{\frac{x_{e}}{x_{e}^{sc}}+\frac{x_{e}^{sc}}{x_{e}}-2\cos^{2}\phi_{e}^{sc}\sin^{2}\theta_{e}^{sc}}}.
\end{equation}
If $\xi_{PD} < PD^{sc}_{e}$, the polarization unit vector of the photon is calculated as \citep{Matt1996,Dreyer_2021} 
\begin{equation}
    \vec{P}^{sc}_{e} = \frac{\brac{\vec{P}_{e} \times \vec{D}^{sc}_{e}} \times \vec{D}^{sc}_{e} }{\left|\vec{P}^{sc}_{e}\right|},
\end{equation}
otherwise, $\vec{P}^{sc}_{e}$ is randomly drawn perpendicular to $\vec{D}^{sc}_{e}$. 

\subsection{Parallelization of the code using Numba CUDA and Cupy}
The algorithm discussed above is a good candidate for parallelization as the evolution and polarization calculations of each photon up until the final binning and summing steps can be treated as entirely independent of each other. 

Parallelization of the code is implemented by employing the use of both the Cupy Python package \citep{cupy2017}, and the Numba just-in-time (jit) compiler developed for Python, leveraging the large number of processors of the GPU to drastically shorten the overall run time of the code. Particularly the Nvidia compute unified device architecture (CUDA) implementations of NUMBA that allow for the use of graphical processing units (GPUs) in general-purpose processing is used \citep{NUMBA,Cuda2008,cuda}. This is done by coding the algorithm as discussed throughout section \ref{sec:MCAlg} and the various deviations discussed in later sections in the Python programming language. The majority of the calculations of section \ref{subsec:overview} is calculated using vectorized implementations of the Cupy package. In cases where vectorized Cupy solutions are not easily implemented such as the numerical calculation of the scattered polar and azimuthal angles, subroutines are decorated in the code with the \textit{@cuda.jit} command that signals to the compiler to run the associated processes on the GPU. Once all of the individual photon calculations are run in parallel on the GPU, the data is saved and a second Python program is run to bin the data and perform the final steps of the algorithm in order to calculate the final $Q_{bin}$ and $U_{bin}$ and the resulting $\Pi_{bin}$ and $\chi_{bin}$ values.

When running processes on the GPU 
through the Numba CUDA interface, care must be taken to treat the local CPU and local memory of the machine and that of the GPU as separate devices \citep{Cuda2008,cuda}. More precisely, all initial values required to run the subroutines must either be copied from local memory to the GPU memory or be initialized directly in GPU memory. All final outputs from the GPU subroutines must be copied back from the GPU memory to the local machine memory. To this extent the Cupy Python package \citep{cupy2017} is used to create and manage all of the required arrays.

Additionally, as the code is primarily a Monte Carlo code and employs the use of many randomly drawn values, the quality of the random number generator (RNG) used is 
important. 
The RNG used by the Numba CUDA compiler is the Xorshiro128+ generator developed by \citet{xorshiro128}. The generator has a period of $2^{128}-1$.



\section{Backscatter-dominated Cork Model}
\FloatBarrier

The Backscatter-dominated cork model by \citet{Vyas1} assumes an instantaneous pulse of radiation originating from the centre of a star intercepted by a cork shaped plasma. The cork is formed due to heated material from the stellar envelope of a collapsing star expanding outward and expelled by jetted material and radiation. This is expected to be 
more applicable to long GRBs.

Radiation produced through pair annihilation in the center of the star then propagates radially outward through the jet funnel interacting with the baryonic cork from the back end. The photons are scattered by the cork in a predominantly backward direction, but will still be observed along the direction of the jet axis due to the relativistic motion of the cork \citep{Vyas1,Vyas2}. 

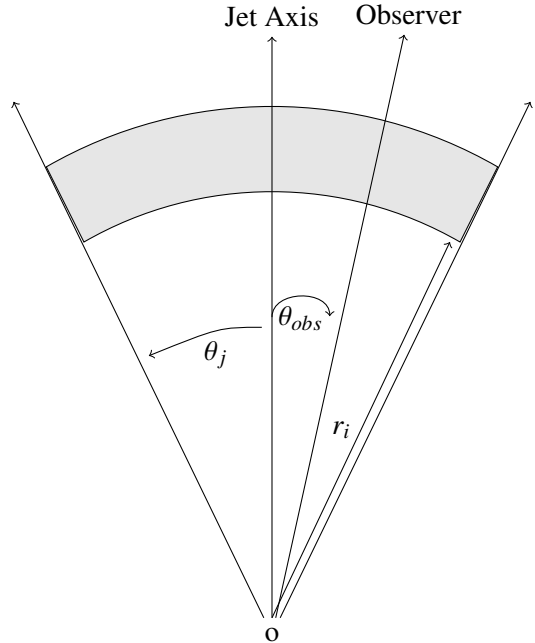
\begin{figure}
    \centering
    \begin{tikzpicture}
        \filldraw[fill=gray!20!white] (0,0) arc (60:120:5cm) node [pos=0.2] (blackBR) {} node [pos=0.3] (blackBL) {} -- (-5.5,1) node [pos=2] (corkLT) {} arc (-240:-300:6cm) node [pos=0.7] (blackTL) {} node [pos=0.8] (blackTR) {} node [pos=0.75] (observer) {}-- cycle node (corkRB) {} node [pos=-1] (corkRT) {}  ;
        \draw node[below] (Origin) at (-2.5,-5cm) {o};
        \draw node (Jet) at (-2.5,3cm) {Jet Axis};
        \draw[->] (Origin) to (corkRT);
        \draw[->] (Origin) to node (corkLAx) {} (corkLT);
        \draw[->] (Origin) to node (JetAx) {} (Jet);
        \draw[->] (Origin.north) to node[left] (r) {$r_{i}$} (corkRB.west);
        \draw[->] (JetAx) .. controls +(west:7mm) .. node[below] {$\theta_{j}$} (corkLAx);
        \draw[draw=none] (Origin) to node [pos=1.2, above] (observerT) {Observer} (observer) ;
        \draw[->] (Origin) to node[left] (obs) {} (observerT) ;
        \draw[->] (JetAx) .. controls +(north:0.5cm) and +(north:0.5cm) .. node[right, below] {$\theta_{obs}$} (obs);
        
    \end{tikzpicture}
    \caption{Diagram showing the geometry of the backscatter-dominated cork model recreated from \citet{Vyas1}. $\theta_{j}$ indicates the cork opening angle, with $r_{i}$ indicating the distance from the star center to the rear edge of the cork.}
    \label{fig:BSDiagram}
\end{figure}

Figure \ref{fig:BSDiagram} shows the geometry of the backscatter-dominated cork model. Consider an expanding cork traveling in the direction of the jet with bulk Lorentz factor $\Gamma$, with a half opening angle $\theta_{j}$, and $\theta_{obs}$ indicating the direction to the observer. The cork is assumed to be hot and optically thick with a co-moving density at radius $r$ of  
\begin{equation}
    n = \dot{M}/\brac{\pi m_{p}\Gamma\beta_{\Gamma}c r^{2}\sin^{2}\brac{\theta_{j}}},
\end{equation}
where $n$ is the co-moving density of the electron plasma that constitutes the cork.
Here $\dot{M}$ is the mass outflow rate, $m_{p}$ is the mass of a proton, $\beta_{\Gamma}$ is the expansion speed in terms of the speed of light $c$. The vertical width of the cork is taken to be $r_{i} / \Gamma$.  

\subsection{Drawing the photons}\label{sec:BSPhot}

For this model photons are drawn from a modified pair annihilation spectrum:
\begin{equation}
    F\brac{\epsilon} = C_{0}e^{-\nicefrac{C_{1}\epsilon^{2}}{\Theta_{r}^{2}}},
\end{equation}
with corresponding probability distribution function (PDF)
\begin{equation}
    p\brac{\epsilon} = \frac{C_{0}}{\epsilon}e^{-\nicefrac{C_{1}\epsilon^{2}}{\Theta_{r}^{2}}},
\end{equation}
where $C_{0} = 2\times10^{40}$,  $C_{1} = 0.045$, and $\Theta_{r} = \nicefrac{kT_{r}}{m_{e}c^{2}}$ \citep{Vyas2}. Additionally, $k$ is the Boltzman constant, $m_{e}$ electron mass, and $T_{r}$ is the temperature of the underlying thermal spectrum. This is done using a rejection technique where $p\brac{\epsilon}$ is estimated by the simpler PDF $g\brac{\epsilon}=\epsilon^{-1}$. A random value $\xi_{s1}$ is drawn and the photon energy calculated as $\epsilon=\epsilon_{min}\brac{\nicefrac{\epsilon_{max}}{\epsilon_{min}}}^{\xi_{s1}}$, with $\epsilon_{min}=10^{-6}$ and $\epsilon_{max}=30$ the range of possible values for $\epsilon$ \citep{Zdziarski_1984}. A secondary random value $\xi_{s2}$ is drawn and compared against the acceptance ratio such that if $\xi_{s2}<e^{-\nicefrac{C_{1}\epsilon^{2}}{\Theta_{r}^{2}}}$, the drawn energy is accepted, otherwise a new $\epsilon$ must be drawn. The direction of the photon $\brac{\theta,\phi}$ is drawn such that $\theta\in\brac{0,\theta_{j}}$ and $\phi\in\brac{0,2\pi}$.

\subsection{Applying the algorithm to the backscatter-dominated cork model}

In order to apply the IC scattering MC algorithm as described in section \ref{sec:MCAlg} to the backscatter-dominated cork model, first a photon is drawn as described in section \ref{sec:BSPhot}, with initial energy $\epsilon_{\rm lab}$, wave four-vector $\underline{k}_{\rm lab}$
and initially placed on the inner surface of the cork at radius $r_{i}$. Following this, the position $\underline{r}_{\rm lab}$ of the photon is calculated in Cartesian coordinates, with the temporal element taken to be zero. An electron is drawn as described in section \ref{subsec:electron} and the photon is transformed into the electron rest frame temporarily in order to calculate $\sigma_{KN}$.

Once back in the lab frame, we draw a random distance along the mean free path of the photon in the direction that the photon is travelling in as
\begin{equation}
    l = -\frac{\brac{\ln\brac{1-\xi_{l}}}}{\sigma_{KN}n}. 
\end{equation}
The position of the photon within the cork is then updated as 
\begin{equation}
    \underline{r}_{\rm lab}^{updated} = \underline{r}_{\rm lab}+\brac{\frac{hc}{\epsilon_{\rm lab}}}\underline{k}_{\rm lab}l.
\end{equation}
Following this, IC scattering is simulated as described in \ref{subsec:overview}. This process is repeated until the photon either leaves the cork or a large enough number of scatterings have occurred so that the photon is considered to be lost in the cork.

If the photon is found to have exited the cork, the linear Stokes parameters $Q$ and $U$ of the photon are calculated as described in section \ref{subsec:overview}.

When binning photons to calculate observed quantities, the geometry of the system needs to be carefully considered.
Under the assumption that the cork is optically thick, if $\theta_{obs}<\theta_{j}$, then there is a region $\theta_{obs}\pm \Gamma_{BL}^{-1}$ from which photons cannot reach the observer.


\subsection{Results}
The code developed by applying the algorithm to the backscatter-dominated cork model is first run for a test case closely matching that presented by \citet{Vyas1}. For this first test case, monoenergetic seed photons with energy $\epsilon_{\rm lab}=0.5 \, \rm MeV$ are injected into a relativistically expanding cork with bulk Lorentz factor $\Gamma = 20$, $\theta_{j}=0.1$ and $\dot{M}= 10^{33} \, \rm g \, s^{-1}$. The inner edge of the cork is initially at radius $r_{i}=10^{12,5}\rm cm$ and the comoving dimensionless electron temperature of the cork is chosen as $\Theta_{e}=0.0169$. $N=9.6\times10^{7}$ photons are used in the simulation, with $\sim 99.2\%$ of the photons backscattered out of the cork within the termination criterion of 25 scattering events and $\sim 33$~\% of the backscattered photons are scattered out of the cork after one scattering event.

\begin{figure}[!h]
    \centering
    \includegraphics[width=0.5\textwidth]{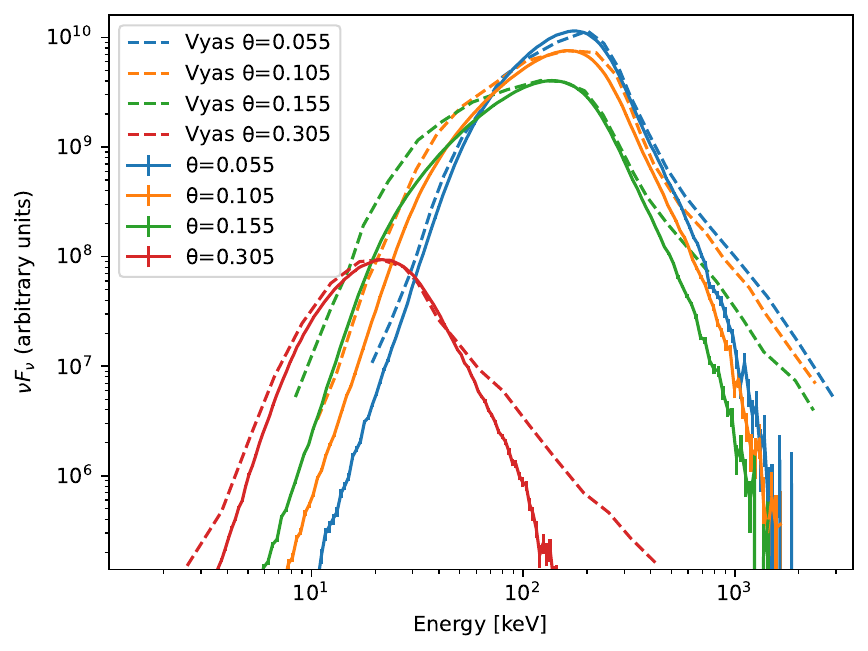}
    \caption{
    SED of the backscatter model for monoenergetic seed photons and viewing angles $\theta_{obs}=0.055$, $0.105$, $0.155$ and $0.305$. The dashed lines show the results obtained by \citet{Vyas1} for the same set of parameters.}
    \label{fig:BS_Mono_nuFnu}
\end{figure}

\begin{figure}[!h]
    \centering
    \includegraphics[width=0.5\textwidth]{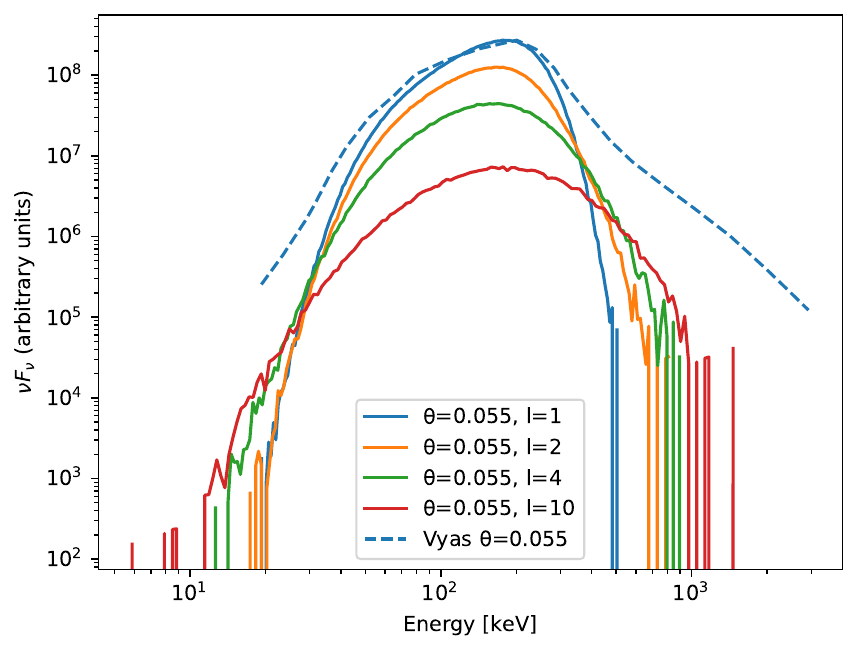}
    \caption{
    SED of the backscatter model for monoenergetic seed photons and viewing angles $\theta_{obs}=0.055$, $0.105$, $0.155$ and $0.305$ resticted to number of scattering events $l$. The dashed lines show the total SED result obtained by \citet{Vyas1} for the same set of parameters.}
    \label{fig:MonoScatterings}
\end{figure}

Figure \ref{fig:BS_Mono_nuFnu} shows the comparison of our results with the results found by \citet{Vyas1}, vertically scaled for comparison. We find good agreement between our results and that of \citet{Vyas1} overall. 
However, in our results, the high-energy tail appears suppressed. 
This is attributable to Klein-Nishina limits at higher energies for photons undergoing multiple scattering events as is illustrated in Figure \ref{fig:MonoScatterings}.
\FloatBarrier

\begin{figure}[!h]
    \centering
    \includegraphics[width=0.5\textwidth]{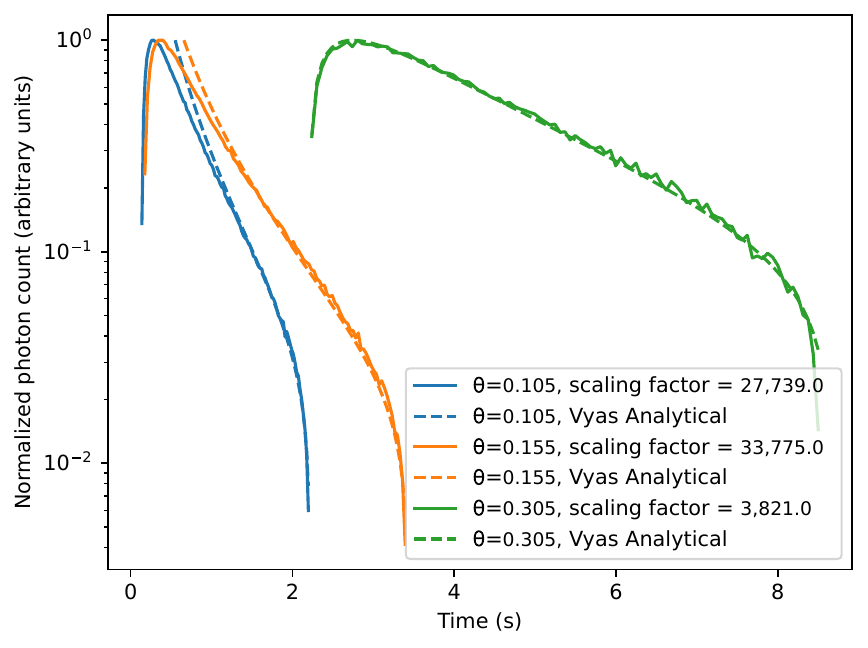}
    \caption{
    Lightcurve of the backscatter model for monoenergetic seed photons and viewing angles $\theta_{obs}=0.055$, $0.105$, $0.155$ and $0.305$. The dashed lines show the theoretical temporal result derived by \citet{Vyas1} for the same set of parameters.}
    \label{fig:BS_Mono_Lc}
\end{figure}

Figure \ref{fig:BS_Mono_Lc} shows the comparison of the light curves obtained from the code to the theoretical light curves derived by \citet{Vyas1} from the geometry of the cork model. As can be seen from the figure our results are in good agreement with the expected theoretical light curve, indicating that the code accurately replicates the backscatter-dominated cork model. As shown by \citet{Vyas1}, the shape of the light curves originates from the geometry of the model, with photons originating from an angular position closest to the viewing angle observed first, with photons originating from positions further away within the cork opening angle arriving at later times. Figures \ref{fig:BS_Mono_nuFnu} and \ref{fig:BS_Mono_Lc} serve to validate the implementation of the backscatter-dominated cork model in our code. It is worth noting that the slight fluctuations visible at the edges of the curves in Figures \ref{fig:BS_Mono_nuFnu} and \ref{fig:MonoScatterings}, and for $\theta_{obs}=0.305$ in Figure \ref{fig:BS_Mono_Lc} are attributable to numerical noise.

\begin{figure}[!h]
    \centering
    \includegraphics[width=0.5\textwidth]{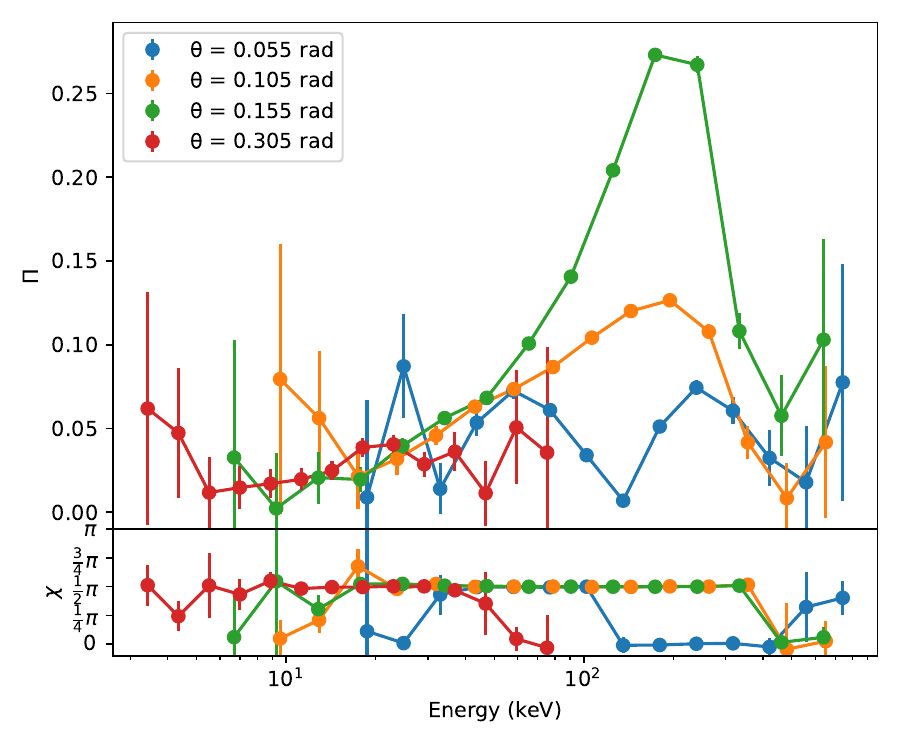}
    \caption{
    Polarization degree and angle as a function of energy for the backscatter model for monoenergetic seed photons and viewing angles $\theta_{obs}=0.055$, $0.105$, $0.155$ and $0.305$. }
    \label{fig:BS_Mono_PolE}
\end{figure}

\FloatBarrier
Figure \ref{fig:BS_Mono_PolE} shows the polarization output of the system as a function of energy for several fixed viewing angles, with the maximum polarization degree ($\Pi$) observed at $\theta_{obs}=\theta_{j}+\Gamma^{-1}$, consistent with theoretical expectations resulting from relativistic aberration. Additionally, the maximal polarization degree for each viewing angle occurs at the peak energy for the relevant viewing angle.   

\begin{figure}[!h]
    \centering
    \includegraphics[width=0.5\textwidth]{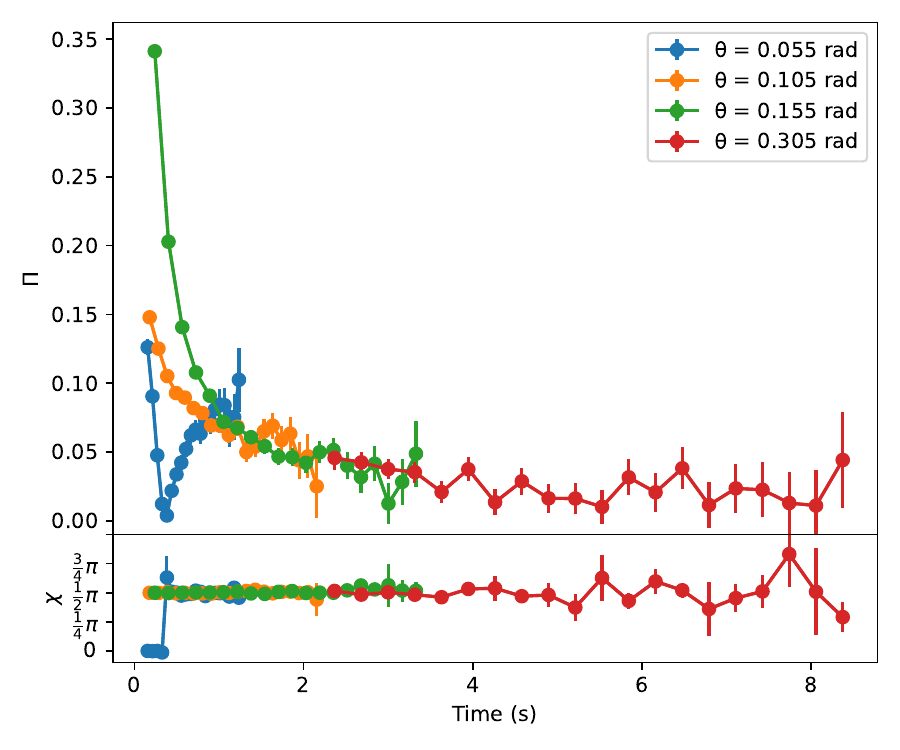}
    \caption{
    Polarization degree and angle as a function of time for the backscatter model for monoenergetic seed photons and viewing angles $\theta_{obs}=0.055$, $0.105$, $0.155$ and $0.305$. }
    \label{fig:BS_Mono_PolT}
\end{figure}

Figure \ref{fig:BS_Mono_PolT} shows the polarization degree and polarization angle ($\chi$) as a function of time for several viewing angles. Off axis viewing angles show an initial maximum polarization that steeply drops off to polarization degrees below $5\%$. This is due to the finite geometry of the cork. As explained by \citet{Vyas1}, the first photons arriving at the observer originate from angular positions closest to $\theta_{obs}$. These photons are more likely to have similar scattered directions in the comoving emission frame, and thus are more coherently aligned resulting in the highest possible polarization degree. As the photons from larger angular positions across the cork are observed, the distribution of possible scattered directions and beamed directions increases resulting in lower observed polarization degrees. 

Both the energy and time resolved polarization results show a polarization angle aligned perpendicular to the jet direction for viewing angles outside the jet cone. This is due to the isotropic distribution of electrons in the comoving emission frame and the higher probability of photons to be forward or backward scattered relative to the electron direction. The result is that a larger fraction of photons are scattered into directions close to $\theta_{s,\rm em}\sim \pi/2$ in the emission frame that are then beamed into the off axis viewing angles. 

Viewing angles within the jet cone show an evolution in the polarization angle initially aligned along the jet axis and then progressing to an alignment perpendicular to the jet axis. This is again due to the scattering geometry of the system. As indicated by \citet{Vyas1}, for viewing angles within the jet cone, a dark region of opening angle $\sim\Gamma^{-1}$ centred around the viewing direction exists for which no photons can reach the observer due to the effect of relativistic aberration. In this case, the first photons to reach the observer are photons immediately outside this dark region. The majority of photons beamed into the viewing angle from the angular position immediately outside of the dark region are scattered out of the cork at angles close to $\theta_{s,\rm em} =\pi$ resulting in a polarization angle aligned with the jet direction \citep{1987BegelmanSikora}. As photons from larger angular positions relative to the viewing direction are observed, the scattering angle in the emission frame shifts to angles closer to $\theta_{s,\rm em}\sim\pi/2$ resulting in the polarization angle shifting towards directions aligned perpendicular to the jet direction. This also explains the results shown in Figure \ref{fig:BS_Mono_PolE}, as the photons scattered into angles close to $\theta_{s,\rm em} =\pi$ are more likely to be head-on.
Coincidentally, the alignment of the polarization angle along the jet direction for observers within the jet cone is also a standard expectation within Compton drag models as the one discussed in section \ref{sec:ComptonDrag} below, albeit without the evolution of the polarization degree and angle as a function of energy and time. This is very briefly touched on by \citet{Lazzati3}. This similarity arises due to the similar scattering geometries within the two models, where in both cases photons reaching the observer at these angles are necessarily strongly backscattered.

\FloatBarrier
As a second validation for the implementation of the backscatter-dominated cork model in our code, we present the results obtained for a scenario similar to that presented by \citet{Vyas2} where a cork of more moderate temperature is assumed and 
the seed photons are assumed to originate from a modified pair annihilation spectrum described in section \ref{sec:BSPhot}.

Setting $\Gamma = 100$, $\theta_{j} = 0.1$, $\Theta_{r}=0.3$, $\Theta_{e}=0.25$, $r_{i}=10^{12.5}$~cm, and $\dot{M}=10^{33}\, g \, s^{-1}$, the code is run for $N=9.6\times10^{7}$ photons, with identical scattering statistics to the mono energetic case described above.

\begin{figure}[h]
    \centering
    \includegraphics[width=0.5\textwidth]{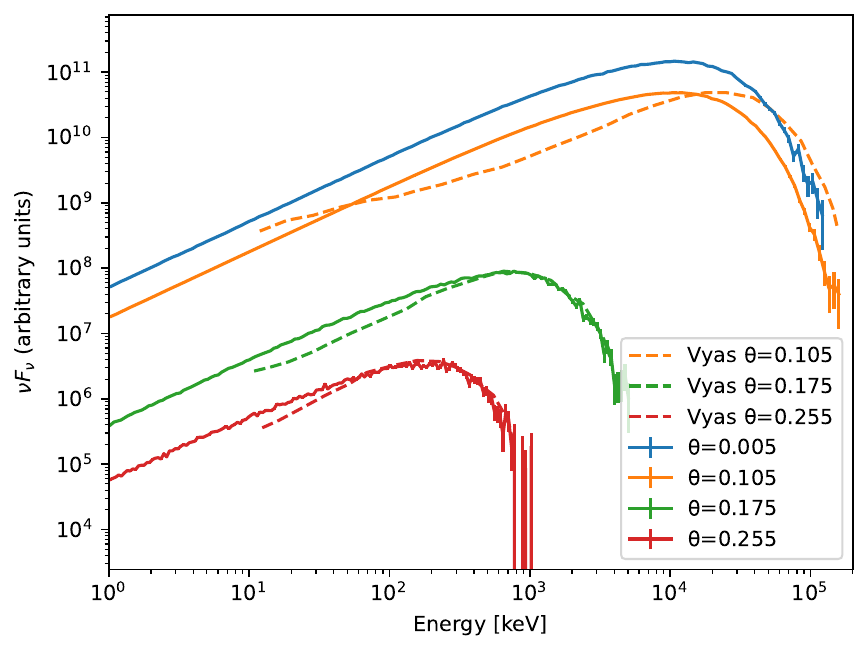}
    \caption{
    SED of the backscatter model for viewing angles $\theta_{obs}=0.005$, $0.105$, $0.175$, and $0.255$. The dashed red line shows the result obtained by \citet{Vyas2} for the same set of parameters.}
    \label{fig:BS_nuFnu}
\end{figure}

\begin{figure}[!h]
    \centering
    \includegraphics[width=0.5\textwidth]{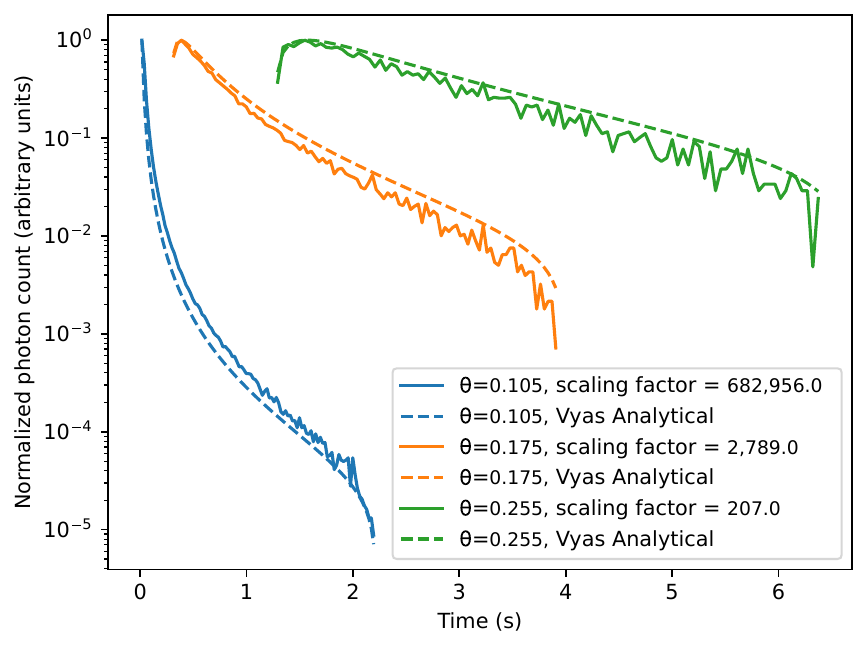}
    \caption{Normalized lightcurve for viewing angles $\theta_{obs}=0.105$, $0.175$, and $0.255$.}
    \label{fig:BS_Lc}
\end{figure}

Figure \ref{fig:BS_nuFnu} shows the SEDs obtained. The results are given for viewing angles $\theta_{obs}=0.105$, $0.175$, and $0.255$. We find that the results from our code and that found by \citet{Vyas2} are in good agreement. Additionally, Figure \ref{fig:BS_Lc} shows normalized simulated lightcurves for the different viewing angles, with the amount of photons received decreasing as $\theta_{obs}$ goes from $0$ to $\pi / 4$. In both figures the minor fluctuations seen in the curves are again attributable to numerical noise, as is evident from the small scaling factor used for $\theta_{obs}=0.255$ in Figure \ref{fig:BS_Lc}, where the fluctuations are the worst.

\begin{figure}[ht]
    \centering
    \includegraphics[width=0.5\textwidth]{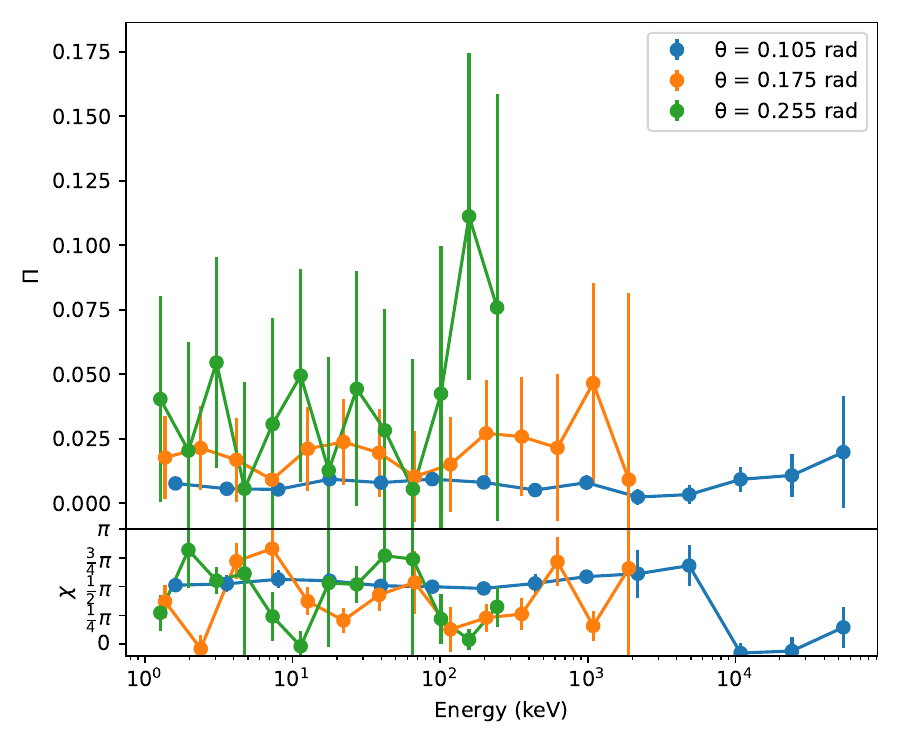}
    \caption{The top panel shows the time-integrated polarization degree and the bottom panel shows the polarization angle as a function of energy for viewing angles $\theta_{obs}=0.105$, $0.175$, and $0.255$.}
    \label{fig:BS_PolE}
\end{figure}

\begin{figure}[hb]
    \centering
    \includegraphics[width=0.5\textwidth]{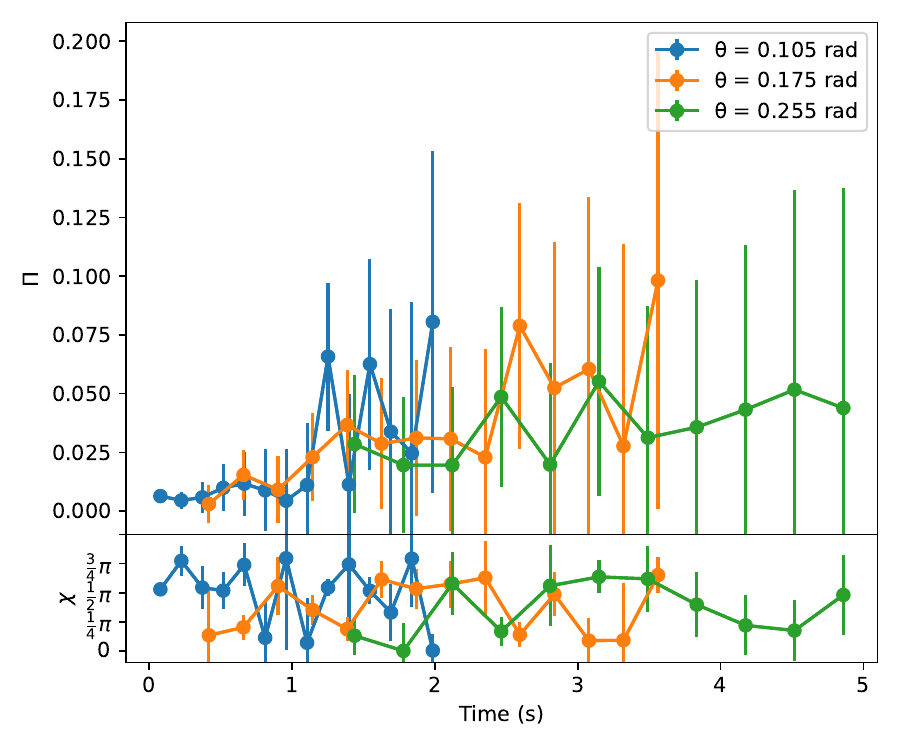}
    \caption{The top panel shows the energy-integrated polarization degree and the bottom panel shows the polarization angle as a function of time for viewing angles $\theta_{obs}=0.105$, $0.175$, and $0.255$.}
    \label{fig:BS_PolT}
\end{figure}

\begin{figure}[!h]
    \centering
    \includegraphics[width=0.5\textwidth]{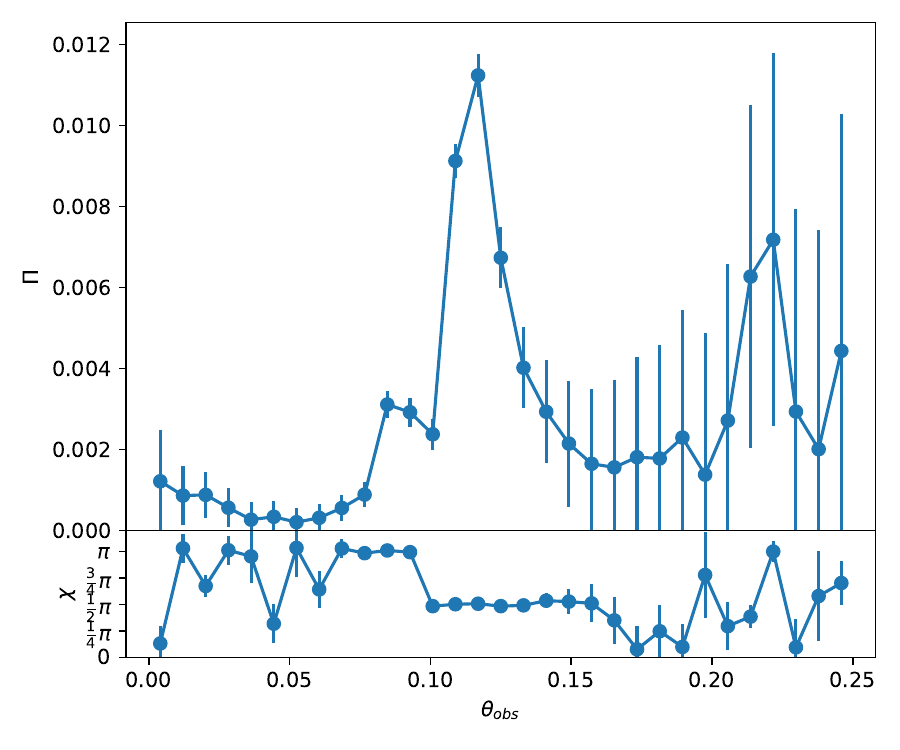}
    \caption{The top panel shows the energy-integrated polarization degree and the bottom panel shows the polarization angle as a function of viewing angle $\theta_{obs}$.}
    \label{fig:BS_PolTheta}
\end{figure}

\FloatBarrier
Figures \ref{fig:BS_PolE} and \ref{fig:BS_PolT} show the resulting polarization degree and polarization angle as a function of energy and time, respectively. In both cases the overall polarization degree is found to be low with a larger uncertainty for larger viewing angles stemming from a lower number of photons observed at larger viewing angles. The maximum polarization degree is expected for viewing angles close to $\theta_{obs}=\theta_{j}+\Gamma^{-1}=0.11$, however, Figure \ref{fig:BS_PolTheta} illustrates that for the chosen parameters, the largest observed polarization degree is only $\sim1\%$.

The largely incoherent polarization angles seen in Figures \ref{fig:BS_PolE} and \ref{fig:BS_PolT} are mainly caused by the low photon counts observed for larger viewing angles as well as the extremely low polarization degree. Figure \ref{fig:BS_PolTheta} does show some consistency with the results found for the mono energetic case where the polarization angle for viewing angles situated within the jet cone are aligned along the jet direction. However, due to the low polarization degree observed, definitive conclusions on this cannot be drawn. 

\FloatBarrier
\section{Compton drag model}\label{sec:ComptonDrag}

In this section we consider the Compton drag model by \citet{Lazzati1}. The model assumes a fireball with bulk Lorentz factor $\Gamma_{BL}\brac{z}$ propagating inside a conical funnel cavity with opening angle $\psi$. The walls of the funnel, with temperature $T (z)$, emit blackbody radiation. The distance $z_{0}$  is assumed to be the distance at the end of the acceleration phase and $z_{*}$ is taken as the radius of the progenitor star. The fireball is also assumed to be cold in the co-moving frame to remain consistent with the underlying assumptions of the model \citep{Lazzati2}. The temperature of the walls of the funnel is parameterized as 
\begin{equation}
    T\brac{z} = T_{0}\brac{\frac{z}{z_{0}}}^{-b} = T_{*}\brac{\frac{z}{z_{*}}}^{-b},
    \label{eq:CD_Temp}
\end{equation}
where $T_{0}=T\brac{z_{0}}$, $T_{*}=T\brac{z_{*}}$, and $b$ is a free parameter. The local radiation density is approximated as $U\brac{z}=aT^{4}\brac{z}$. At the distance 
\begin{equation}
    z_{T} = \brac{\frac{\sigma_{T}E_{f}}{\pi\psi^{2}m_{p}c^{2}\Gamma_{0}}}^{\nicefrac{1}{2}} \sim 3.7\times 10^{14}\psi^{-1}_{-1}E_{f,51}^{\nicefrac{1}{2}}\Gamma^{\nicefrac{1}{2}}_{0,2} cm
\end{equation}
the fireball becomes optically thin \citep{Lazzati2}. Here, $E_{f}$ is the energy of the fireball, $m_{p}$ is the mass of a proton, and $\Gamma_{0}$ is the initial bulk Lorentz factor of the fireball.

\citet{Lazzati2} find the following solution for the bulk Lorentz factor as a function of distance:
\begin{equation}
\begin{split}
    \Gamma_{BL}\brac{z} =& \Gamma_{0}\brac{1+2\pi\psi^{2}aT_{0}^{4}\Gamma_{0}^{2}z_{0}^{3}\frac{\bracc{\brac{\nicefrac{z}{z_{0}}}^{3-4b} -1 }}{\bracc{E_{f}\brac{3-4b}}}}^{-1}\\
    & \text{for  } z_{0} < z < z_{*},\\
\text{and}&\\
    \Gamma_{BL}\brac{z} =& \Gamma_{*}\brac{1+2\pi\psi^{2}aT_{*}^{4}\Gamma_{0}\Gamma_{*}z_{*}^{3}\frac{\bracc{\brac{\nicefrac{z}{z_{*}}}^{3-g} -1 }}{\bracc{E_{f}\brac{3-g}}}}^{-1}\\
    & \text{for  } z_{*} < z < z_{T},\\
    \label{eq:CD_BL}
\end{split}
\end{equation}
where $g$ is a free parameter, and $b$ is the same as in equation \ref{eq:CD_Temp}. 

\subsection{Calculating initial conditions for the Compton drag model}\label{subsec:CDInitial}

Letting $z_{i}$ be the initial position along the jet axis of a target photon for Compton scattering, the initial energy of the photon $\epsilon$ is calculated from the Plank number spectrum using the temperature $T (z)$ calculated from equation \ref{eq:CD_Temp}. Rewriting equation \ref{eq:CD_BL} to find the normalized speed $\beta_{BL}\brac{z_{i}}$ of the emission region, and integrating this from $z_{0}$ to $z_{i}$ we find the time of the scattering event to be  
\begin{equation}
\begin{split}
    t_{i} & = \frac{1}{c\Gamma_{A}}\int_{z_{0}}^{z_{i}}\bracc{\Gamma_{A}^{2}-\brac{1+\frac{A}{B}\bracc{\brac{\frac{z}{z_{A}}}^{B}-1}^{2}}}^{-2}dz,
\end{split}
\end{equation}
where $A=2\pi\psi^{2}aT_{0}^{4}E_{f}^{-1}\Gamma_{0}^{2}z_{0}^{3}$, $\Gamma_{A}=\Gamma_{0}$, $z_{A}=z_{0}$, and $B=3-4b$ for $z_{0}<z<z_{*}$, and $A=2\pi\psi^{2}aT_{*}^{4}E_{f}^{-1}\Gamma_{0}\Gamma_{*}z_{*}^{3}$, $\Gamma_{A}=\Gamma_{*}$, $z_{A}=z_{*}$, and $B=3-g$ for $z_{*}\leq z<z_{T}$. Finally, the initial position four-vector is then $\underline{r}_{lab}=\brac{t_{i},r\xi_{r}\cos\brac{\pi\xi_{\theta}},r\xi_{r}\sin\brac{\pi\xi_{\theta}},z_{i}}$,where $r$ is a free parameter. Here $z_{i}$ is chosen such that the number of photons $N_{ph}$ evolved through the system is proportional to $z^{-1}$, with $z_{i}=\nicefrac{z_{0}z_{m}}{\brac{z_{m}+\brac{z_{0}-z_{m}}\xi_{z}}}$.

The initial direction of the photon $\vec{D}^{lab}$ is drawn such that $\theta_{lab} \in \brac{0,\pi}$ and $\phi_{lab} \in \brac{0,2\pi}$.

The electrons are again drawn as in section \ref{subsec:electron}, where we assume $T_{e}\sim 1 \rm keV$ to be consistent with the assumption of a cold fireball.

\subsection{Applying the algorithm to the Compton drag model}

For the case of the Compton drag model, we only consider single IC scattering. In order to determine if a drawn photon undergoes a scattering event a random number $\xi_{\sigma}$ is drawn from a uniform distribution and compared to the ratio $\nicefrac{\sigma_{KN}}{\sigma_{T}}$. If $\xi_{\sigma} < \nicefrac{\sigma_{KN}}{\sigma_{T}}$, Compton scattering is simulated. Beyond this alteration no additional modifications to the algorithm are made when applying this to the Compton drag model. The initial position and energy of the seed photons are calculated as described in section \ref{subsec:CDInitial}. The bulk Lorentz factor of the fireball at the time and position of each scattering event is calculated from equation \ref{eq:CD_BL}. 

\subsection{Results}

\begin{figure}[h]
    \centering
    \includegraphics[width=0.6\textwidth]{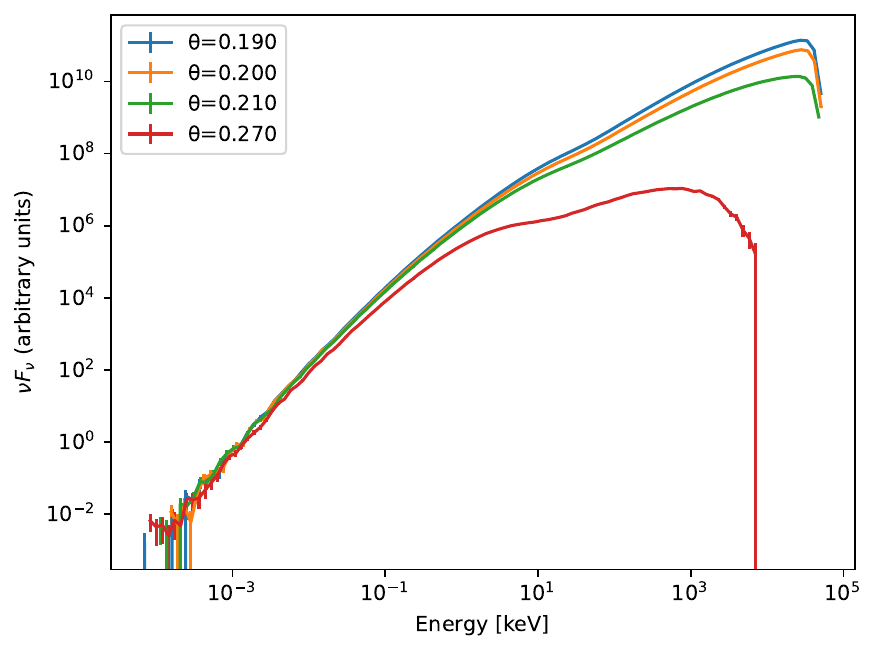}
    \caption{
    Time-integrated SEDs of the Compton drag model for viewing angles $\theta_{obs}=0.19$, $0.2$, $0.21$, and $0.27$.}
    \label{fig:CD_nuFnu}
\end{figure}

Applying the algorithm to the Compton drag model, we set $b=0.5$, $g=2$, $E_{f}=10^{51} \, \rm erg$, $\Gamma_{0}=100$, $\psi=0.2$, $T_{*}=3\times10^{5}$~K, $z_{0}=10^{8}$~cm, and $z_{*}=10^{13}$~cm. We also set the half-opening angle of the jet $\theta_{j}=\psi=0.2$.


\begin{figure}[h]
    \centering
    \includegraphics[width=0.6\textwidth]{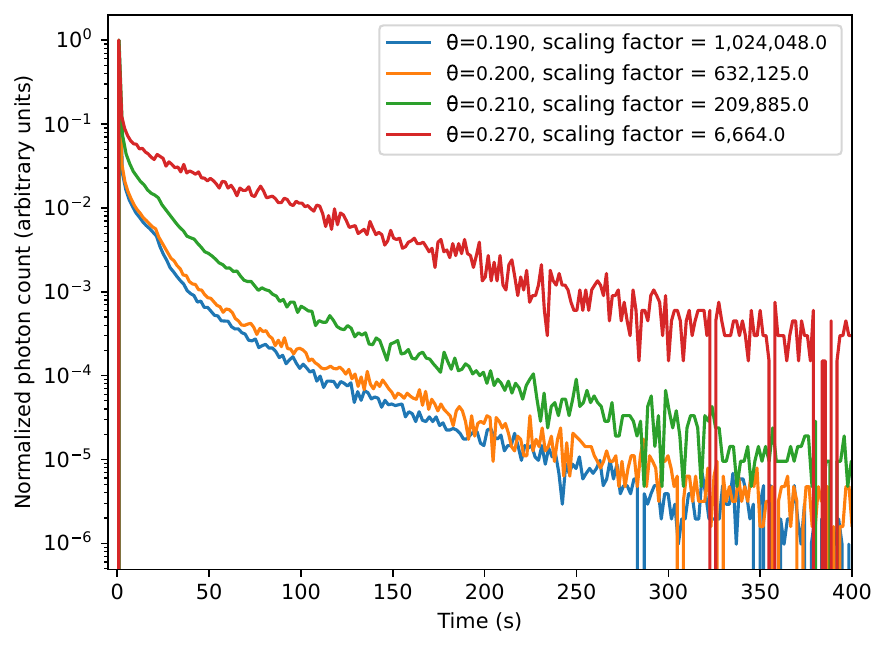}
    \caption{Normalized, energy-integrated lightcurves for viewing angles $\theta_{obs}=0.19$, $0.2$, $0.21$, and $0.27$.
    Here the scaling factor used to normalize the lightcurve is indicated.}
    \label{fig:CD_Lc}
\end{figure}

Figure \ref{fig:CD_nuFnu} shows the SEDs obtained for the given parameters, showing the results for viewing angles $\theta_{obs}=0.19$, $0.2$, $0.21$, and $0.27$.
Figure \ref{fig:CD_Lc} shows the normalized simulated lightcurves.
For all four viewing angles, the lightcurve indicates a large spike of detected photons that exponentially decays. In Figure \ref{fig:CD_nuFnu} there is again minor numerical noise present in the low energy tail. In Figure \ref{fig:CD_Lc} the majority of the minor fluctuations are attributable to numerical noise. As the amount of photons received decays with time, the noise increases.


\begin{figure}[h]
    \centering
    \includegraphics[width=0.6\textwidth]{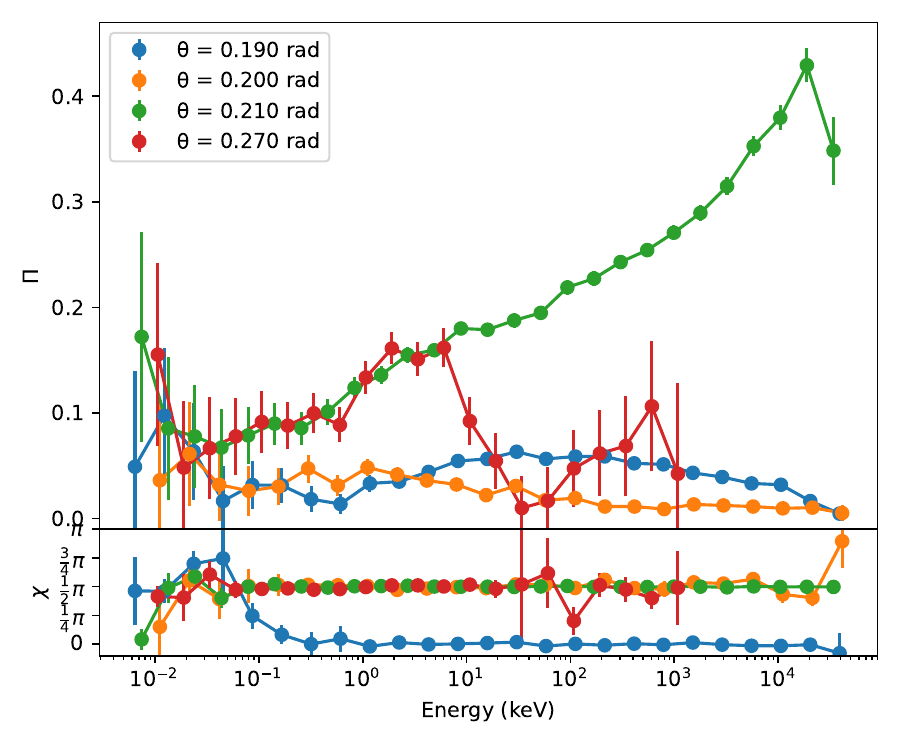}
    \caption{The top panel shows the time-integrated polarization degree and the bottom panel shows the polarization angle as a function of energy for viewing angles $\theta_{obs}=0.19$, $0.2$, $0.21$, and $0.27$.}
    \label{fig:CD_PolE}
\end{figure}

\begin{figure}[h]
    \centering
    \includegraphics[width=0.6\textwidth]{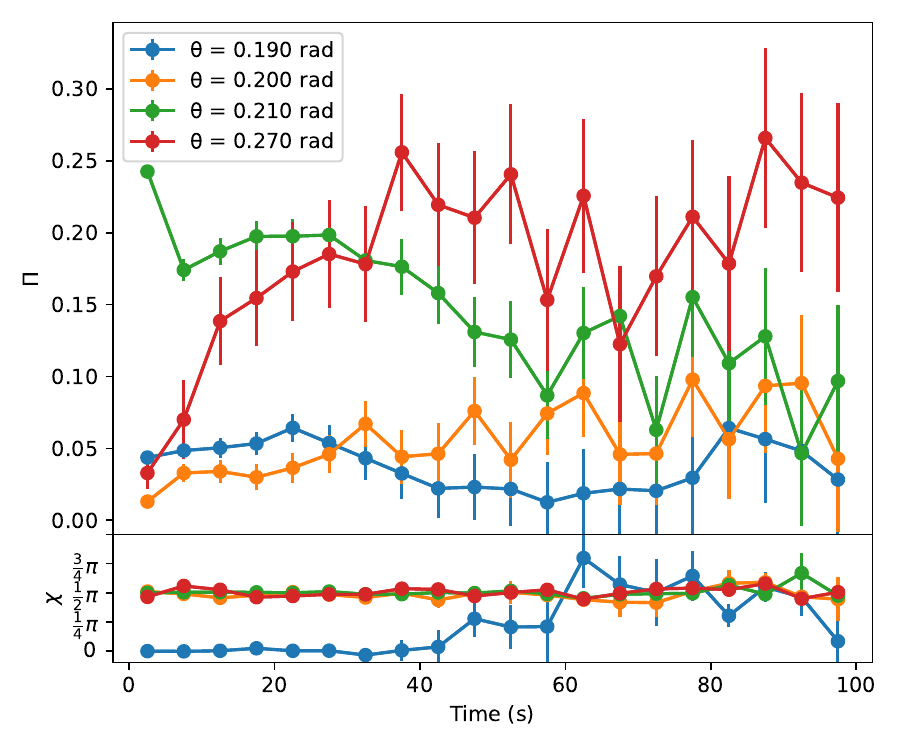}
    \caption{The top panel shows the energy-integrated polarization degree and the bottom panel shows the polarization angle as a function of time for viewing angles $\theta_{obs}=0.19$, $0.2$, $0.21$, and $0.27$.}
    \label{fig:CD_PolT}
\end{figure}

Figures \ref{fig:CD_PolE} and \ref{fig:CD_PolT} show the polarization degree ($\Pi$) and polarization angle ($\chi$) as a function of scattered photon energy and time, respectively. In both cases the polarization degree 
remains flat below $10\%$ for viewing angles $\theta_{obs}\leq\theta_{j}$, where we again see a polarization angle aligned parallel to the projection of the jet direction on the sky for $\theta_{obs}=0.19$ within the jet cone arising for the same reason as discussed previously for the mono energetic back scatter dominated model. As expected the results show a scattering angle constant at $\chi \sim \pi/2$ for $\theta_{\rm obs}\geq\theta_{j}$. 

For $\theta_{obs}=0.21$ we see a maximal amount of polarization, consistent with photons scattering into $\theta_{\rm em}^{\rm sc}=\pi/2$ in the emission frame, with the maximum polarization expected for $\theta_{\rm obs}\sim\theta_{j}+\Gamma_{0}=0.21$. The energy resolved results for $\theta_{obs}=0.21$ indicates a low polarization degree at lower energies that gradually increases to the maximum observed polarization degree of $\Pi\sim 40\%$ at high energies. The time resolved results for $\theta_{obs}=0.21$ indicates an initial maximum amount of polarization that decreases with time. For $\theta_{\rm obs}=0.27$, however, we see that the maximum polarization degree of $\Pi\sim18\%$ is only observed for moderate to low energies. Interestingly, the time resolved polarization result for $\theta_{obs}=0.27$ shows an initial low polarization degree that increases with time, although these results may be somewhat contaminated by low number of photon statistics available at $\theta_{obs}=0.27$, particularly for late times, as is evident from Figure \ref{fig:CD_Lc}. From these results a clear picture emerges. As described in Section \ref{subsec:CDInitial}, the largest amount of available seed photons with the highest energies are present deep within the stellar funnel at $z\sim z_{0}$ where the fireball propagating through the funnel has the largest bulk Lorentz factor $\Gamma=\Gamma_{0}$. This results in the polarization degree observed for $\theta_{obs}=0.21$ at high energies. As the fireball propagates through the funnel, it is slowed down due to the Compton drag effect, resulting in a decrease in $\Gamma$, described by equation \ref{eq:CD_BL}. This means the angle where the maximum amount of polarization is expected to be observed increases to larger viewing angles at later times, resulting in the decrease of observed polarization degree at $\theta_{obs}=0.21$ and an increase in the observed polarization degree at $\theta_{obs}=0.27$ with time. The decreasing $\Gamma$ and lower temperatures of the funnel wall at later times in the fireball propagation then naturally also accounts for the maximum polarization degree for $\theta_{obs}=0.27$ seen at lower energies than for $\theta_{obs}=0.21$.

\begin{figure}[h]
    \centering
    \includegraphics[width=0.6\textwidth]{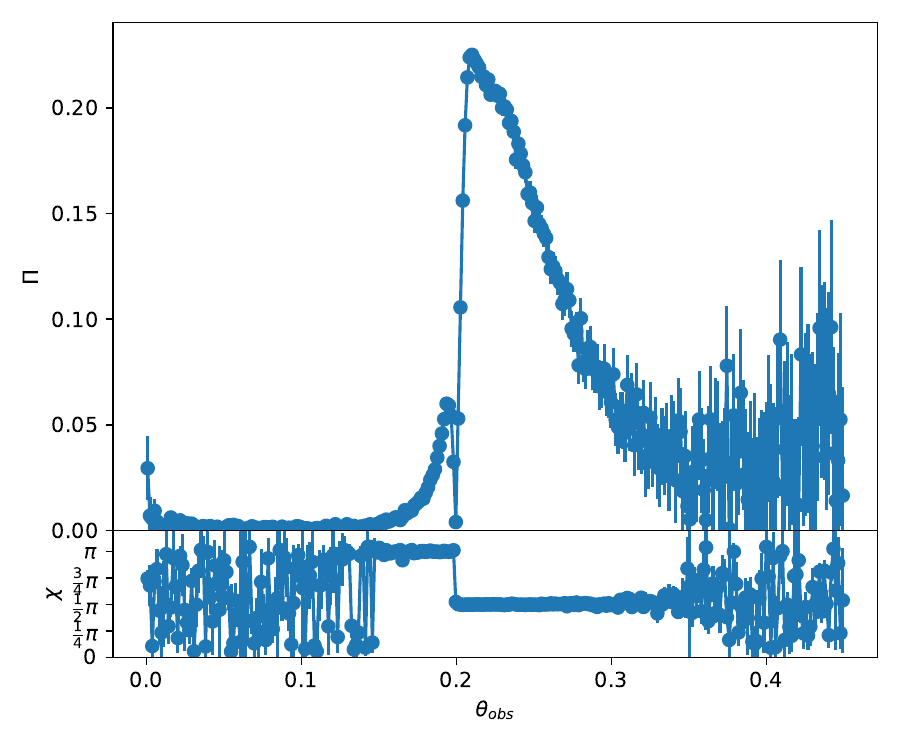}
    \caption{The top panel and bottom panels show polarization degree and polarization angle respectively as a function of viewing angles.}
    \label{fig:CD_PolTheta}
\end{figure}

As a validation for the polarization results obtained from our implementation of the Compton drag model, Figure \ref{fig:CD_PolTheta} shows the time and energy integrated polarization degree and angle as a function of viewing angle. From the figure we see that the maximum amount of polarization expected for the chosen parameters occurs at $\theta_{\rm obs}\sim\theta_{j}+\Gamma_{0}^{-1}$. Additionally Figure \ref{fig:CD_PolTheta} shows that the observed polarization angle for $\theta_{\rm obs}<\theta_{j}$ is aligned parallel to the jet axis projected onto the sky. This is in good agreement with similar results found by \citet{Lazzati3}, and \citet{Toma_2009}.

\section{Conclusions}
Applying the IC scattering polarization algorithm to two different IC scattering models for GRBs, we find that the overall polarization predictions both for the case of the backscattered cork model \citep{Vyas1} and the Compton drag model \citep{Lazzati1} at on axis observation angles fall below $10\%$. The Compton drag model at $\theta_{obs} \sim \theta_{j} + 1 / \Gamma$ is found to result in large polarization degrees in agreement with \citet{Lazzati3}. 

For the case of the backscattering cork model, the polarization is found to be largely independent of energy when considering a realistic pair annihilation spectrum of seed photons and a electron plasma of moderate temperature. The polarization degree as a function of time reveals some dependence on time which can be attributed to the geometry of the model. 

Applying the algorithm to the Compton drag model, we find that the simulated lightcurves show the bulk of the detected photons in a sharp initial peak that decays over time consistent with the application of the Compton drag model to long GRBs. The resulting polarization degree is found to be low and mostly constant in both time and energy for on axis observing angles. However, for larger observing angle, the polarization shows some evolution likely attributed to the geometry and dynamics of the system. Investigating the polarization result for the Compton drag model, we find the results consistent with the results of \citet{Toma_2009} for the chosen set of parameters and viewing angles. For $\theta_{obs} \sim \theta_{j}+1 / \Gamma$, however, we find a notably larger polarization degree, 
as expected, since in this case, target photons are scattered by $\sim 90^o$ in the co-moving frame, which optimizes Compton-induced polarization.

Our findings of a low polarization degree for both the back-scattering-dominated cork model for realistic seed photons and the Compton drag model for on axis observers are consistent with POLAR observations of low polarization degrees in the $50-500$~keV range \citep{Polar}. It is worth noting here that we do see temporal evolution of the polarization angle as hinted at by the POLAR observations and explored by \citet{Parsotan_2022} and \citet{Ito_2024} however, our results are not well aligned to these results. Future prospects to compare our findings with GRB polarization observations include the planned 2027 launch of the Compton Spectrometer and Imager (COSI) which will operate in the $0.2-5$~MeV range \citep{COSI}.

Comparing our polarization results to synchrotron dominated models such as work done by \citet{Lan_2020} and \citet{Lan_2021}, we find, for the case of the backscatter dominated cork model, a significantly lower polarization degree, and a similar polarization degree for the used Compton drag model. The overall structure of the temporal evolution of the polarization degree also differs significantly.

In comparison with other photospheric models, such as work done by \citet{Ito_2024} and \citet{Parsotan_2022}, we find some agreement in the prediction of the polarization degree for both the backscatter dominated cork model, and the Compton drag model. We also find some evidence for an evolving polarization angle for on axis viewing angles in agreement with the work done by those authors.

\begin{acknowledgments}
This work is based on the research supported in part by the National Research Foundation of South Africa (Grant Numbers: 144799) and acknowledge that opinions, findings and conclusions or recommendations expressed in any publication generated by the NRF supported research is that of the author(s) alone, and that the NRF accepts no liability whatsoever in this regard.

Furthermore, this work was partially supported through the South African Gamma-Ray Astronomy Programme (SA-GAMMA), funded by the South African Department of Science, Technology, and Innovation through the National Research Foundation..
\end{acknowledgments}

\begin{contribution}

\end{contribution}

\bibliography{Investigating_polarization_signatures_from_GRB_models}{}
\bibliographystyle{aasjournalv7}

\end{document}